\documentclass[11pt,preprintnumbers,aps,amssymb,nofootinbib,amsmath]
{revtex4}
\usepackage{epsfig,epsf}
\usepackage{bm} 
\usepackage{txfonts} 
\newcommand{\beq}{\begin{equation}}
\newcommand{\beql}[1]{\begin{equation}\label{#1}}
\newcommand{\eeq}{\end{equation}}
\newcommand{\bea}{\begin{eqnarray}}
\newcommand{\eea}{\end{eqnarray}}
\newcommand{\eq}[1]{(\ref{#1})}
\newcommand{\fig}[1]{Fig.~\ref{#1}}
\renewcommand{\sec}[1]{Sec.~\ref{#1}}
\newcounter{topiccounter}
\renewcommand{\b}[1]{{\bf #1}} 

\newcommand{\as}{\alpha_s}
\newcommand{\bas}{\bar\alpha_s}

\newcommand{\aver}[1]{\left\langle #1 \right\rangle}
\newcommand{\Q}{\mathcal{Q}}
\newcommand{\dip}{q\bar q}
\newcommand{\xP}{x_{I\!\!P}}
\begin{document}

\title{Diffractive hadron production in DIS off heavy nuclei and gluon saturation}

\author{Kirill Tuchin and  Dajing Wu\\}

\affiliation{
Department of Physics and Astronomy, Iowa State University, Ames, IA 50011}

\date{\today}


\begin{abstract}
We calculate the cross section for  diffractive hadron production in deep inelastic scattering off heavy nuclei in the framework of gluon saturation/color glass condensate. We analyze the  kinematic region of the future Electron-Ion Collider. 
We argue that coherent and incoherent diffractive channels are very sensitive to the structure of the nuclear matter at low $x$. This expresses itself in a characteristic dependence of the cross sections on rapidity and transverse momentum of the produced hadron and on the nuclear weight. We also discuss dependence on the scattering angle and argue that both coherent and incoherent cross sections may be within experimental reach at EIC.

\end{abstract}

\maketitle

\section{Introduction}\label{sec:intr}

Diffraction is one of the most effective tools for investigating the structure of the nuclear matter at low values of Bjorken variable $x$. Its hallmark is  large rapidity gaps (LRG) in  rapidity distribution of the produced hadrons. At high energies, these gaps correspond to scattering processes mediated by exchange of a collective gluon state with vacuum quantum numbers, known as Pomeron. On the other hand, according to the Pomerantchuk theorem, high energy asymptotic of QCD is driven by the Pomeron exchange (see e.g.\ \cite{Levin:1997mg}). Hence, measurements  of diffractive structure functions at HERA attracted a lot of interest. Indeed, diffractive physics at HERA yielded many exciting results that heralded the dawn of the new QCD regime of  gluon saturation/color glass condensate (CGC) \cite{Gribov:1983tu,Mueller:1986wy,McLerran:1993ni,Kovchegov:1996ty,Balitsky:1995ub,JalilianMarian:1997jx,Jalilian-Marian:1997dw,Kovner:2000pt,Iancu:2000hn,Ferreiro:2001qy}.  

A possible launch of Electron Ion Collider (EIC) will open new avenues in studying the physics of diffraction in high energy nuclear physics. It will not only allow probing lower $x$ and  measure dependence of diffractive processes on nuclear weight, but also make possible studying less inclusive processes. One such process, diffractive hadron production in DIS  is the subject of this paper.   Our goal is to make predictions for DIS on a nucleus  at the EIC kinematic region based on the CGC theory. We argue that diffractive hadron production is very sensitive to parameters of CGC and thus can be  very effective instrument in extracting properties of the nuclear matter at low $x$.   Gluon saturation effects on diffractive gluon production in DIS on proton at HERA have been discussed in \cite{Bartels:1998ea,Gotsman:1996ix,GolecBiernat:1998js,GolecBiernat:1999qd,Gotsman:2000gb,Munier:2003zb,Marquet:2004xa,GolecBiernat:2005fe,Marquet:2007nf}. A concise discussion of the gluon saturation effects in semi-inclusive DIS on nuclei is given in \cite{Kowalski:2007rw,Kowalski:2008sa,Kugeratski:2005ck}.

This article is structured as follows. In \sec{sec:diffr} we review the formalism developed in our previous publications \cite{Li:2008jz,Li:2008se,Tuchin:2008np}, which allows to calculate coherent and incoherent diffractive gluon production in the regime of coherent scattering $l_c\gg R_A$, where $l_c= 1/(M_Px)$ is the coherence length in the nucleus rest frame. Coherent diffractive gluon production is the process $\gamma^*+A\to X+h+[LRG]+A$. The corresponding cross section is given by  Eqs.~\eq{xsect}--\eq{w-f} and \eq{cd10} below. For heavy nuclei $A^{1/3}\sim 1/\as^2\gg 1$ and at high energies this type of diffractive process dominates over the incoherent diffraction, which is the process $\gamma^*+A\to X+h+[LRG]+A$ with $A^*$ being excited nucleus. Nevertheless, at EIC energies, cross sections for coherent and incoherent diffraction processes are  often comparable \cite{Tuchin:2008np}. In $pA$ collisions their dependences on gluon rapidity $y$ and transverse momentum $\b k$ and on atomic weight $A$ are quite different. Therefore, as was pointed  
out in \cite{Tuchin:2008np}, it is important to separately  measure the  contributions 
of these diffractive processes. In \sec{appA} we calculate these contributions using the b-CGC model \cite{Kowalski:2006hc} for the color dipole scattering amplitude. As in \cite{Li:2008se} we characterize the nuclear effect using the nuclear modification factor (NMF) for diffractive processes defined in \eq{nmf}. The results of our numerical calculations are presented in \fig{main-figure}. The most interesting features of  the NMF's are (i) strong dependence of coherent diffractive NMF on gluon rapidity $y$ (or $\xP$); (ii) near independence of incoherent diffractive NMF on $y$ and (iii) independence of both NMF's on the photon virtuality. This results are discussed in detail in \sec{appA}. 

Separation of coherent and incoherent diffractive contributions pose a great experimental challenge because it requires measurements of very small scattering angles $\theta = 2\sqrt{-t/W^2}$, where $t$ is the moment transfer and $W$ is the center-of-mass energy per nucleon of $\gamma^*A$ process. We address this problem in \sec{sec:t-dep}. Dependence of the coherent cross section on momentum transfer $t$ is given by \eq{coh-t}. It is seen that it decreases as $1/|t|^3$ at $|t|\gg 1/R_A^2$, where $R_A$ is the nuclear radius. On the other hand, incoherent diffraction cross section decreases exponentially as $e^{-|t|R_p^2/4}$, but at much larger momentum transfers $t>1/R_p^{2}$ as seen in \eq{idp5}. The results of the calculation are plotted in \fig{fig:t-dep}. As expected coherent diffraction dominates at small momentum transfers $-t$ while the incoherent one at large $-t$. However, due to different functional form of $t$-dependences, the two contributions become of the same order at about $-t\sim R_P^{-2}$ and remain comparable even at larger momentum transferes. The corresponding scattering angle for $W= 100$~GeV is $\theta\approx  0.13^\text{o}$ and is very weakly dependent on the hadron transverse momentum, $\xP$ and photon virtuality $Q^2$.
It seems that such scattering angles are  within the experimental reach and hopefully the two contribution can be separated.

\section{Diffractive gluon production}\label{sec:diffr}

\subsection{Dipole cross section}

\begin{figure}[ht]
      \includegraphics[height=4cm]{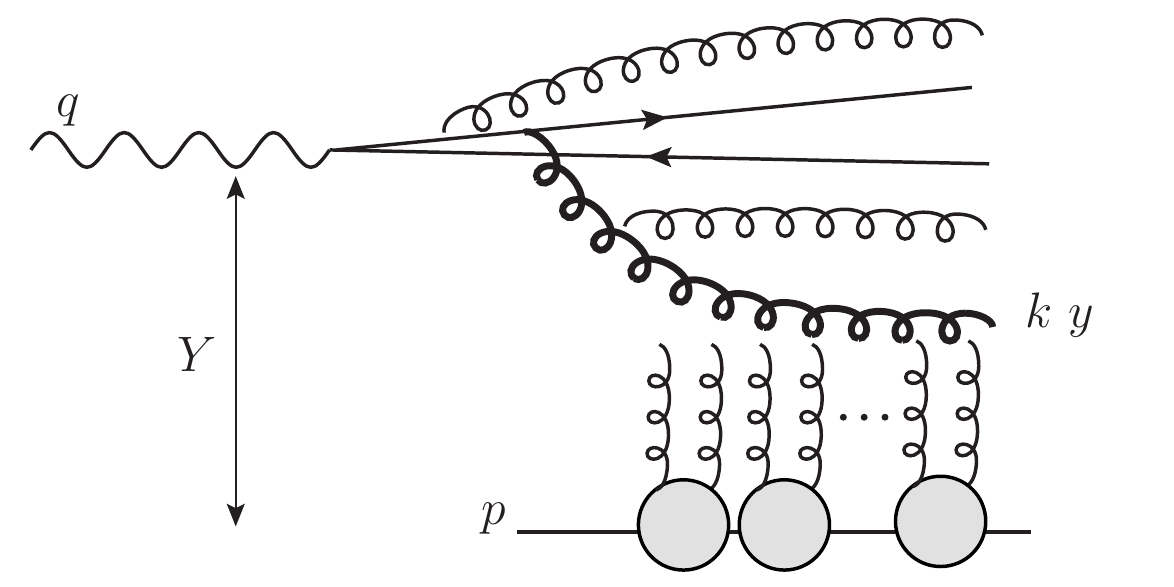} 
  \caption{One of the diagrams contributing to the diffractive production of a gluon with transverse momentum $\b k$ and rapidity $y$. $y$ is also the rapidity gap of the process. Unconnected $t$-channel gluons indicate all possible attachments to the $s$-channel gluons, quark and anti-quark.   }
\label{diagram}
\end{figure}

Consider diffractive production of a gluon of transverse momentum $\b k$ at rapidity $y$. Let the total rapidity interval be $Y=\ln (1/x)$, where $x= Q^2/W^2$, $Q^2$ is photon virtuality and $W$ the center-of-mass energy of $\gamma^*N$ scattering. 
Cross section for diffractive gluon production reads \cite{Li:2008bm} 
\beql{xsect}
\frac{d\sigma_\text{diff}^{\gamma^*A}(Q^2,x, k,y)}{d^2 k\, dy}
=\int \frac{d^2 r}{2\pi^2} dz \,\Phi^{\gamma^*}(Q, r, z) \,
\frac{d\sigma_\text{diff}^{q\bar qA}( r,x, k,y)}{d^2 k\, dy}\,,
\eeq
where 
\beql{dip.xs}
\frac{d\sigma_\text{diff}^{q\bar qA}( r,x, k,y)}{d^2 k\, dy}
\eeq
is the differential cross section for the diffractive gluon production by a  $q\bar q$ dipole (a.k.a.\ onium) of transverse size $r$ scattering off a nucleus. Eq.~\eq{xsect} generalizes the quasi-classical result derived in \cite{Kovchegov:2001ni,Kovner:2001vi,Kovner:2006ge}.
Other kinematic variables that are often used are $\beta$ and $\xP$. They are defined as  $\ln(1/\beta)=Y-y $ and $\ln(1/\xP)=y$, where  $Y-y$ is the rapidity interval between the photon and the produced gluon. We work in the approximation $\as \ln(1/x)\sim 1$, $\as\ln(1/\beta)\sim 1$. Diffractive production in the region $\beta\lesssim 1$ was addressed in \cite{Gotsman:1996ix,Wusthoff:1997fz}.  We assume that the produced gluon is at the edge  of the rapidity gap, so that the total rapidity gap in the process is $y$, see \fig{diagram}.

Virtual photon light-cone wave-function appearing in \eq{xsect} reads
\begin{eqnarray}\label{w-f}
\Phi^{\gamma^*}(Q, r,z)&=& \Phi^{\gamma^*}_T(Q, r,z)+\Phi^{\gamma^*}_L(Q, r,z)\\
\Phi^{\gamma^*}_T(Q, r,z)&=&2N_c\sum_f\frac{\alpha_\text{em}^f}{\pi}\{a^2 K_1^2( r a)[z^2+(1-z)^2]+m_f^2 K_0^2( r a)\}\\
\Phi^{\gamma^*}_L(Q, r,z)&=&2N_c\sum_f\frac{\alpha_\text{em}^f}{\pi} 4Q^2z^2(1-z)^2K_0^2( r a)
\end{eqnarray}
where $a^2= Q^2z(1-z)+m_f^2$, $\alpha_\text{em}^f= e^2 z_f^2/(4\pi)$, with $z_f$ being electric charge of quark $f$.  Subscripts $L$ and $T$ refer to the longitudinal and transverse polarizations respectively. 

\subsection{Coherent and incoherent diffraction}

We will consider two types of diffractive processes on nuclei -- coherent and incoherent diffraction. Coherent diffraction is a process in which nucleus stays intact.  This  corresponds to elastic dipole scattering. At very high energies, such processes constitute half of the total dipole--nucleus cross section, another half being the inelastic processes. Therefore, contribution of  coherent diffractive processes is expected to rise with energy. Unfortunately, experimental observation of coherent  diffraction is challenging because it requires measurements at very small scattering angles, i.e.\ at very small momentum transfers $|t|\sim 1/R_A^2$. We discuss this in detail in \sec{sec:t-dep}. 

Another type of diffractive process is incoherent diffraction when the nucleus decays into colorless remnants. This process occurs at the nuclear edge  where partial scattering amplitude at a given impact parameter is less than unity. Share of this contribution in the total inelastic cross section decreases with energy and with nuclear weight. Importance of incoherent diffraction stems from the fact that it measures fluctuations of the color glass condensate near its quasi-classical mean-field value. Typical momentum transfer in this case is $|t|\sim 1/R_p^2$, i.e.\ determined by the inverse width of the diffuse region; it is much larger than in the case of coherent diffraction, which allows easier experimental study. In this section we discuss coherent and incoherent diffraction separately, assuming no experimental cuts on the minimal scattering angle.  

Cross section for coherent diffractive gluon production including the low-$x$ evolution was derived in \cite{Li:2008bm,Li:2008jz} and can be written as
\beq\label{cd10}
\frac{d\sigma_\mathrm{cd}(\b r,x,\b k,y)}{d^2\b k\, dy}=\frac{\as C_F}{\pi^2}\, \frac{1}{(2\pi)^2}
\int d^2b  \int d^2r'\, n_p(\b r, \b r', Y-y)\, |\b I_\mathrm{cd}(\b r',x,\b k,y,\b b)|^2\,,
\eeq
where we introduced an auxiliary transverse vector
\bea\label{icd}
\b I_\mathrm{cd}(\b x- \b y,x,\b k,y,\b b)&=& \int d^2 z \left(\frac{\b z-\b x}{|\b z-\b x|^2}-
 \frac{\b z-\b y}{|\b z-\b y|^2}\right)\,e^{-i\b k \cdot \b z}\,\nonumber\\
 && 
 \times\,
 \bigg\{ 
 -N_A(\b z-\b x,\b b, y) -N_A(\b z-\b y,\b b, y)  +N_A(\b x-\b y,\b b, y)  \nonumber\\
 &&
 +N_A(\b z-\b x,\b b, y) \,N_A(\b z-\b y,\b b, y)
  \bigg\}\,.
\eea
In \cite{Li:2008bm,Li:2008jz} we presented a detailed analytical and numerical analysis of the the coherent diffractive gluon production and discussed applications to $pA$ scattering in \cite{Li:2008se}. 
Similarly, for incoherent diffraction \cite{Tuchin:2008np}
\beq\label{all2}
\frac{d\sigma_\mathrm{id}( r,x, k,y)}{d^2k\, dy}=\frac{\as C_F}{\pi^2}\, \frac{\pi R_p^2}{2(2\pi)^2}
\int d^2b  \int d^2r'\, n(r,  r', Y-y)\, \rho\, T_A(\b b)\, |\b I_\mathrm{ID}( r',x, k,y, b)|^2\,,
\eeq
where 
\bea\label{iid}
&&\b I_\mathrm{id}(\b x- \b y,x,k,y,b)= \int d^2 z \left(\frac{\b z-\b x}{|\b z-\b x|^2}-
 \frac{\b z-\b y}{|\b z-\b y|^2}\right)\,e^{-i\b k \cdot \b z}\,\nonumber\\
 && 
 \times\,
 \bigg\{ 
 \left[ 1-N_A(\b z-\b x,\b b, y) \right]\, \left[ 1-N_A(\b z-\b y,\b b, y) \right]
 \, \left[ N_p(\b z-\b x,0, y) +  N_p(\b z-\b y,0, y)  \right]
 \nonumber\\
 &&
 -   \left[ 1-N_A(\b x-\b y,\b b, y) \right] N_p(\b x-\b y,0, y) 
 \bigg\}\,.
\eea
For numerical calculation we evaluate both vector functions $\b I_\mathrm{id}$ and $\b I_\mathrm{cd}$ in the logarithmic approximation \cite{Li:2008bm,Li:2008jz,Li:2008se,Tuchin:2008np}.  Dipole density $n( r, r', Y-y)d^2r'$ in \eq{all2} is the number of daughter dipoles of size $r'$ produced by a parent dipole of size $r$ in the two-dimensional element of area $d^2r'$  at relative rapidity $Y-y$.  In the diffusion approximation to the leading order BFKL equation \cite{Kuraev:1977fs,Balitsky:1978ic} it is given by: 
\beq\label{ndiff}
n( r, r', Y-y)=\frac{1}{2\pi^2}\frac{1}{rr'}\sqrt{\frac{\pi}{14\zeta(3)\bas\, (Y-y)}}\, e^{(\alpha_P-1)(Y-y)}\, e^{-\frac{\ln^2\frac{r}{r'}}{14\zeta(3)\bas\, (Y-y)}}\,.
\eeq
 
As discussed in detail in \cite{Li:2008se}, nuclear modification factor $R_{AB}$ for coherent diffractive gluon production in the quasi-classical regime (i.e.\ without low-$x$ evolution)  is suppressed for large nuclei and large dipoles as $R_{\dip+ A}\sim A^{1/3}\exp\{-r^2 Q_s^2/4\}$ (modulo logs) for dipole--nucleus scattering. Effect of quantum evolution is twofold. The  larger is the rapidity of the produced gluon $y$, the stronger is the coherence effect that slows down growth of the diffractive $q\bar q+A$ cross section with energy as compared to the diffractive $q\bar q+p$  cross section. As a result, the nuclear modification factor gets an additional suppression in the $\gamma^*$ fragmentation region (forward rapidity). On the other hand, at large $Y-y$, the dipole density \eq{ndiff} in the virtual photon $\gamma^*$ spreads to a wider range of sizes $r'$. Apparently, dipoles with sizes $r'\ll 2/Q_s$  are not suppressed at all. This effect leads to relative enhancement of the nuclear modification factor in the backward versus forward rapidity. A quantitative study of diffractive hadron production requires numerical calculations that we discuss in the next section.

\section{Numerical calculations}\label{appA}

A convenient way to express the nuclear effect on diffractive scattering is to introduce the nuclear modification factor as a ratio of the diffractive cross sections in DIS on a nucleus per nucleon and on a proton \cite{Li:2008se}:
\beql{nmf}
R_\text{cd/id}= \frac{\frac{d\sigma_\text{cd/id}^{\gamma^*A} (Q^2,x, k,y)}{d^2 k\, dy}}{A\,\frac{d\sigma_\text{cd/id}^{\gamma^*p} (Q^2,x, k,y)}{d^2k\, dy}}\,.
\eeq
Cross sections appearing in \eq{nmf} are partonic cross sections \eq{cd10} and \eq{all2}
 convoluted with the LO pion fragmentation function given in \cite{Kniehl:2000hk}. 
 
We performed  numerical calculations  with  b-CGC model of the dipole scattering amplitude $N$ \cite{Kowalski:2006hc}, albeit with a modification: we treat  nuclear and proton profiles as step-functions; the saturation scales are assumed to scale with $A$ as $Q_s^2\propto A^{1/3}$. The advantage of this model is that (i) its form complies with the known analytical approximations to the BK equation and (ii) its parameters are fitted to the low $x$ DIS data. The explicit form of the scattering amplitude $N$   is given by
\beql{kmw}
N(\b r, 0, y)= \,\left\{
\begin{array}{cc}
\mathcal{N}_0\left( \frac{r^2\Q_s^2}{4}\right)^\gamma\,,&\quad r\Q_s\le 2;\\
1-\exp[-a\ln^2(br\Q_s)]\,,&\quad r\Q_s\ge 2\,,
\end{array}\right.
\eeq
where $\Q_s^2$ is the the \emph{quark} saturation scale related to the \emph{gluon} saturation scale $Q_s^2$ -- which we have referred to simply as the `saturation scale' throughout the paper -- by $\Q_s^2= (4/9)Q_s^2$. Its functional form is
\beql{sat.scale}
\Q_s^2= A^{1/3} x_0^\lambda\, e^{\lambda y}\,\mathrm{GeV}^2\,,
\eeq
 The anomalous dimension is 
\beql{anom.dim}
\gamma = \gamma_s+\frac{1}{\kappa\, \lambda \,y}\ln \left( \frac{2}{r\Q_s}\right)\,.
\eeq
Parameters $\gamma_s=0.628$ and $\kappa=9.9$ follow from the BFKL dynamics \cite{Iancu:2002tr}, while $\mathcal{N}_0=0.7$, $x_0=3\cdot 10^{-4}$ and $\lambda=0.28$ are fitted to the DIS data. Constants $a$ and $b$ are uniquely fixed from by the requirement of continuity of the amplitude and its first derivative.

\begin{figure}[ht]
\begin{tabular}{cc}
      \includegraphics[height=5cm]{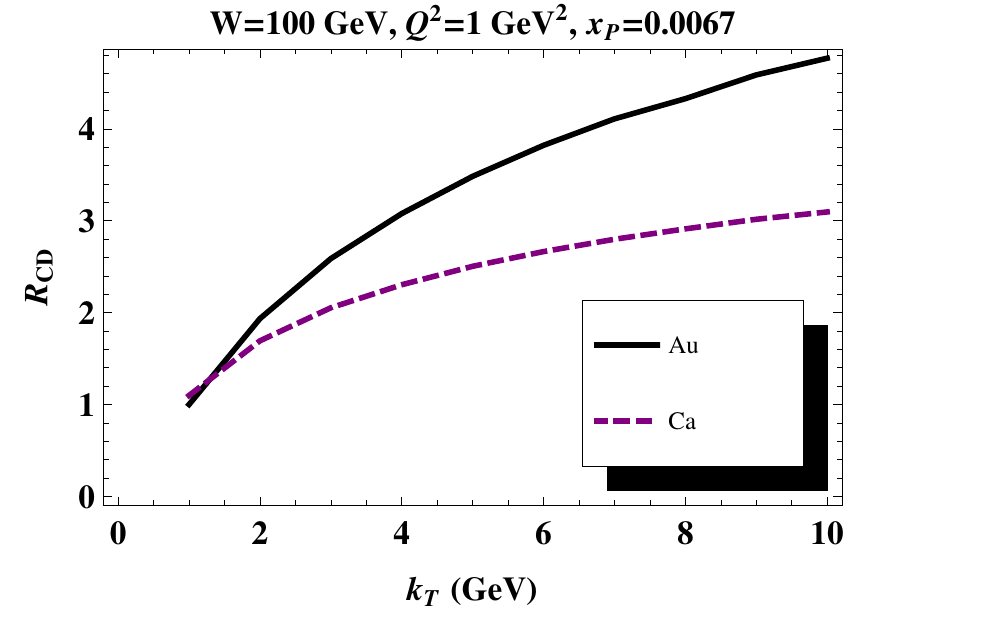} &
      \includegraphics[height=5cm]{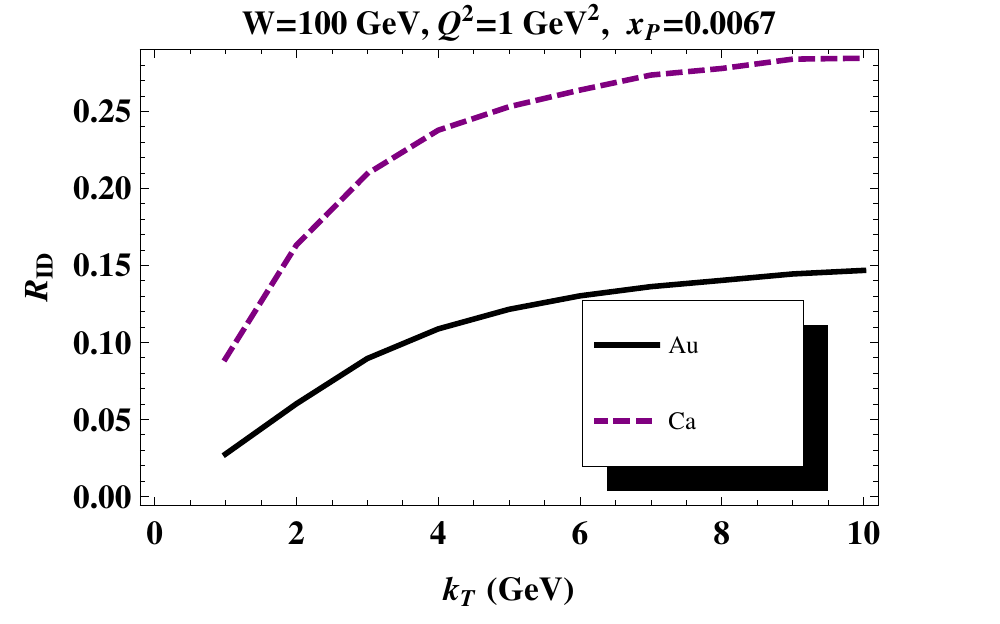}\\[-2mm]
      $(a)$ & $(b)$ \\[2mm]
        \includegraphics[height=5cm]{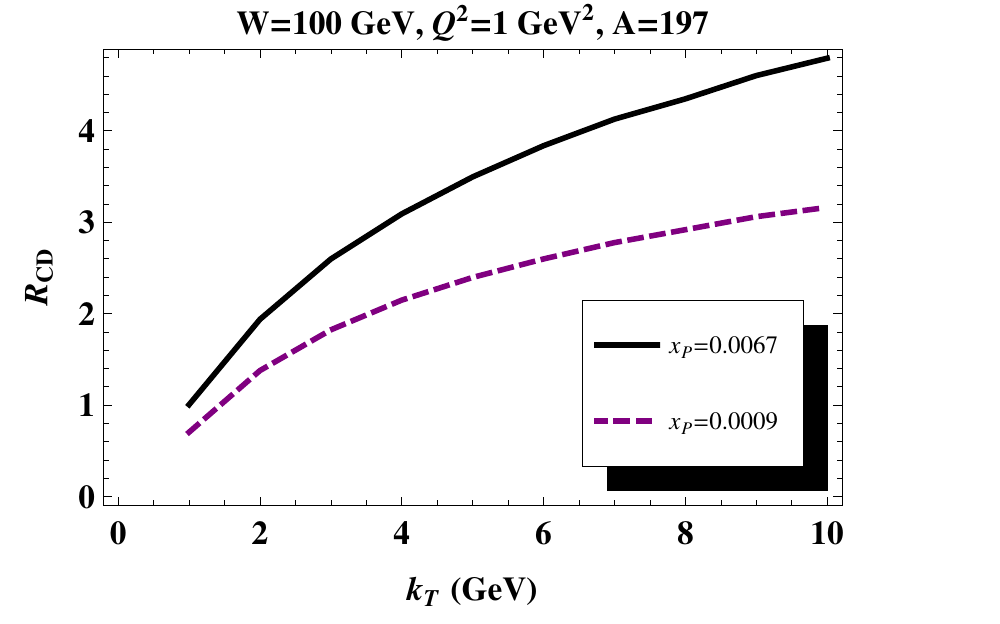} &
      \includegraphics[height=5cm]{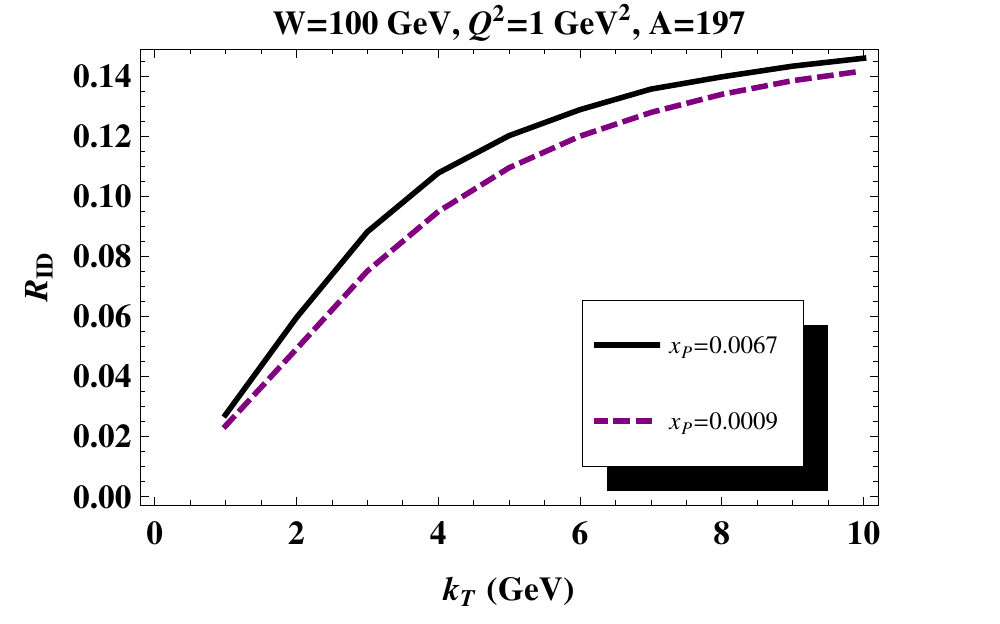}\\[-2mm]
            $(c)$ & $(d)$ \\[2mm]
              \includegraphics[height=5cm]{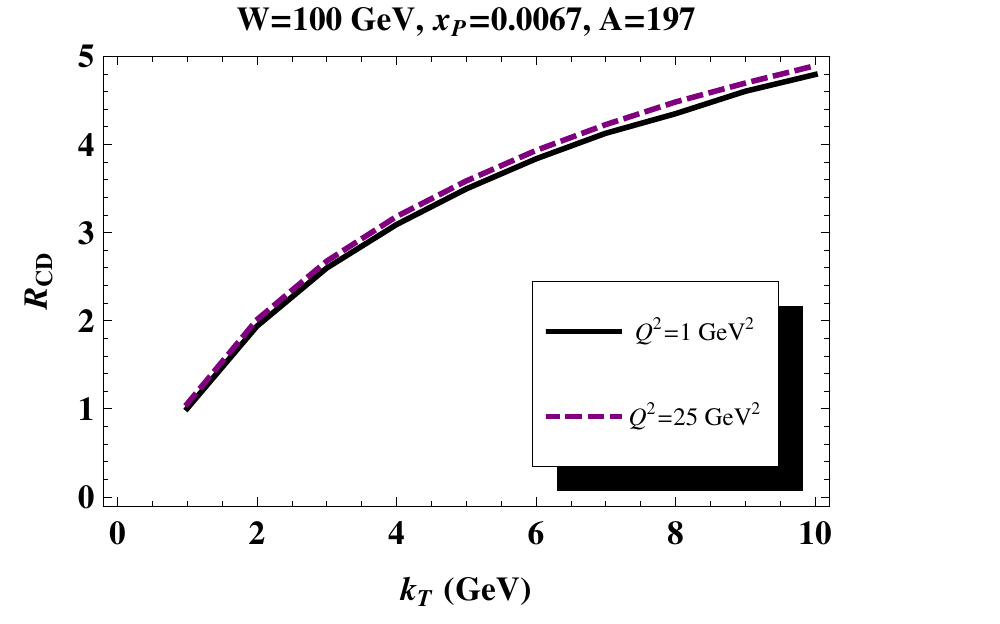} &
      \includegraphics[height=5cm]{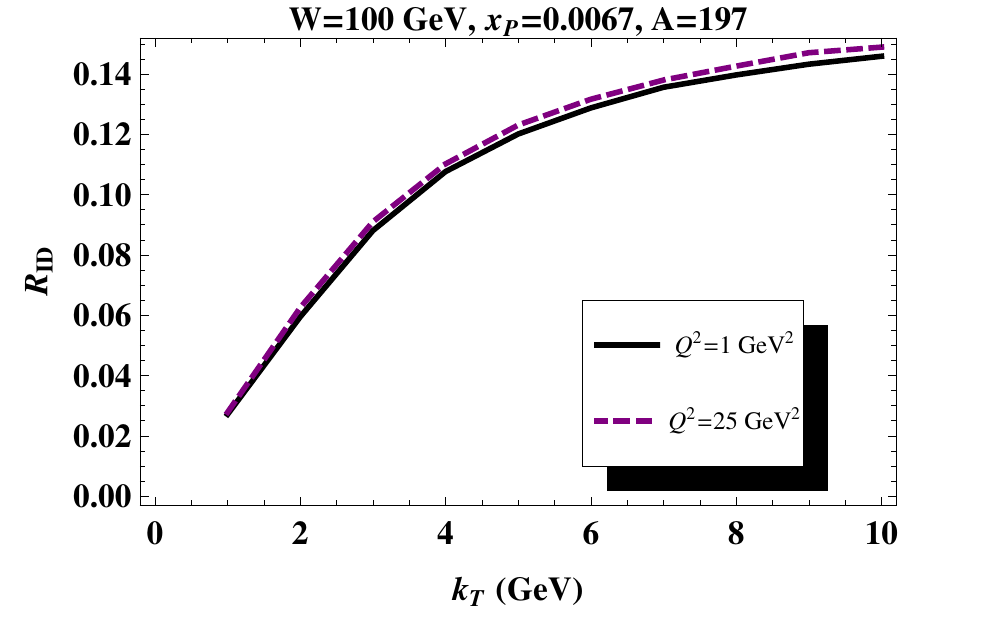}\\[-2mm]
            $(e)$ & $(f)$
      \end{tabular}
  \caption{Nuclear modification factors for coherent (left column) and incoherent (right column) diffractive hadron production at $W=100$~GeV  as a function of the hadron transverse momentum $k_\bot$.  Shown are dependences on: (a),(b) atomic number $A$, (c),(d) hadron rapidity $y$  and (e),(f) photon virtuality $Q^2$. }
\label{main-figure}
\end{figure}

Our results are presented in \fig{main-figure} which exhibits dependence of the nuclear modification factor for coherent (left column) and incoherent (right column) hadron production on transverse momentum  $\b k$. We assumed that the center-of-mass energy of the $\gamma^*A$ collision is $W=100$~GeV per proton, which corresponds to the total rapidity interval $Y=9.2$.  

In \fig{main-figure}~(a,b) we show variation of  the nuclear modification factor with the nuclear weight. We observe that  $R_\text{cd}$ increases with $A$. This is a signature behavior of higher twist effects and, in particular, coherent diffraction. In view of the discussion at the end of the previous section, we infer that the effective dipole size $r'$ produced in the dipole evolution is $r'\ll 2/Q_s$, for otherwise the cross section would decrease for heavier nuclei. As one can see in 
\fig{main-figure}~(e,f), NMF has  no significant $Q^2$ dependence, and hence no $r$ dependence as well. Therefore, even at higher $y$, where evolution effects in the nucleus as well as lack of evolution in $\gamma^*$ could have produced suppression of $R_\text{cd}$ with $A$, no such suppression is observed.  We checked this statement up to the most forward direction allowed by our model $\beta=0.1$. 
  $R_\text{id}$ decreases with $A$ already at midrapidity $y=5$ because the general property of  incoherent diffraction  
is that it vanishes in the limit $A\to \infty$ when all partial amplitudes turn black. 
 
 Rapidity dependence is displayed in \fig{main-figure}~(c,d).  $R_\text{cd}$ rapidly decreases in the forward direction, which is a cumulative effect of evolution in the nucleus and in the virtual photon, whereas  
 $R_\text{id}$ is essentially rapidity independent.  This effect has already been noticed by one of us   in $pA$ case \cite{Tuchin:2008np}. 
It arises because of different physical origins of the two diffractive processes. Coherent 
 diffraction corresponds to elastic scattering of a color dipole on a nucleus, whereas incoherent diffraction is a part of inelastic scattering that originates from the nuclear periphery due to variation of the nuclear density with impact parameter. At low $x$ central impact parameters of a heavy nucleus are black for a typical dipole. Therefore, scattering amplitude of dipole on a heavy nucleus is very different from an incoherent  superstition of dipole-nucleon scattering amplitudes, hence strong variation of the nuclear modification factor with energy/rapidity. On the other hand, incoherent diffraction is non-zero only in the range of impact parameters comparable with the proton radius. Therefore, energy/rapidity dependence of dipole-nucleus and dipole-proton cross section is similar, though the geometry is quite different.

 Finally,   \fig{main-figure}~(e,f) exhibits dependence on photon virtuality $Q^2$, or perhaps better to say no dependence at all. This can be interpreted as  insensitivity of the diffractive cross sections to the size of the parent dipole $r$. Indeed,  as explained in \cite{Li:2008jz},  at $k_\bot \gg Q_s,Q$ diffractive spectra depend only on $k_\bot$. For example, cross section for coherent diffractive gluon production in the asymptotic kinematic region $Q_s\ll 1/r\ll k$ reads (in the double-logarithmic approximation)
 \beq\label{assy}
 \frac{d\sigma^{q\bar qA}}{d^2k\, dy}= \frac{\as C_FS_A}{\pi^{5/2}k^2}N^2(1/k,b,y)\frac{1}{(2\bar \as(Y-y)\ln(rk))^{1/4}}e^{2\sqrt{2\bas (Y-y)\ln (rk)}}\,.
 \eeq
Clearly, $r$-dependence cancels out of the nuclear modification factor. Notice, however, that the EIC kinematic region can hardly be classified as asymptotic, and one would expect large corrections to \eq{assy}. In fact, it is known that corrections to the double-logarithmic approximation are phenomenologically significant (see e.g.\ \cite{Gotsman:2002zi,Gotsman:2000gb}). However, our numerical calculations imply that they cancel in this particular case.  Unfortunately, we are not able to extend this analysis to higher $Q^2$'s without transgressing the region of  applicability of our model.  It would be interesting to analytically investigate the origin of this cancelation.

 \section{t-dependence}\label{sec:t-dep}

 In this section we consider dependence of different diffraction channels on momentum transfer $t$. $t$-dependence translates into dependence on the scattering angle $\theta$. While the dominant contribution to the diffractive cross sections stems from scattering at small angles, only angles larger than some cutoff angle are experimentally accessible. In this section,  we would like to investigate whether  separation of coherent and incoherent contributions is experimentally feasible at EIC.

\subsection{Coherent diffraction}

Consider  dipole--nucleus elastic scattering amplitude $\Gamma^{\dip+ A}(s, \b b, \{\b b_a\})$, where $\b b$ is the dipole impact parameter and $\b b_a$'s are positions of nucleons in the nucleus. Average over the nucleon positions will be denoted as   $\aver{\Gamma^{dA}(s,\b b)}$.  Cross section for elastic dipole scattering is  
\beql{cdx}
\sigma_\text{cd}^{\dip+A}=\int d^2b\, \left| \aver{\Gamma^{\dip +A}(s,\b b)}\right|^2\,.
\eeq
In this representation, \eq{cdx}  is also the coherent diffraction cross section. Fourier image of the dipole-nucleus elastic scattering amplitude carries information about the transferred momentum $\Delta$ ($t= -\b \Delta^2$):
\beq\label{image}
 \aver{\Gamma^{\dip+A}(s,\b \Delta)}= 2\int d^2b \aver{\Gamma^{\dip+A}(s,\b b)} e^{i\b b\cdot \b \Delta}\,.
\eeq
If only two-body forces are taken into account in the scattering amplitude, which amounts  
to neglecting correlations between nucleons, then we can express the scattering amplitude on a nucleus  through the scattering amplitudes on individual nucleons as
\beq\label{gamma.pa}
\Gamma^{\dip+A}(s,\b b, \{ \b b_a\})= 1-\prod _{a=1}^A\left(
1-\Gamma^{\dip+N}(s,\b b-\b b_a)\right)\,.
\eeq
In this approximation, averaging can be performed as 
\beq\label{aver}
\langle \ldots \rangle = \prod_{a=1}^{A}\int d^2b_a \int_{-\infty}^\infty dz \,\rho_A(\b b_a, z) \ldots = \prod_{a=1}^{A}\int d^2b_a \rho T_A(\b b_a)  \ldots
\eeq
 where  $\rho_A(\b b, z) $ is the nuclear density at a given point in the nucleus and $\rho$ is its average over the nucleus volume. 
 
Impact parameter profile of the dipole-nucleon amplitude is traditionally parameterized as 
\beql{dN}
\Gamma^{\dip+N}(s,\b b)= \frac{1}{2}\sigma_\text{tot}^{\dip+N}(s)\,\frac{1}{\pi R_p^2}e^{-b^2/R_p^2}\,,
\eeq
where we neglected a small imaginary part of $\Gamma^{\dip+N}(s,\b b)$. In a heavy nucleus of radius $R_A\gg R_p$, nucleon can be approximated by the delta function in impact parameter space. Thus,
\beql{conv}
\int d^2b_a \Gamma^{\dip+N}(s,\b b-\b b_a) \, \rho T_A(\b b_a)\approx \rho \Gamma^{\dip+N}(s,0) \, \rho T_A(\b b) \,.
 \eeq
Using \eq{aver},\eq{dN},\eq{conv} in \eq{gamma.pa} we derive for heavy nuclei
\beql{hn}
\aver{\Gamma^{\dip+A}(s,\b b)} = 1-e^{-\frac{1}{2} \sigma_\text{tot}^{\dip+N}(s)\rho T_A(b)}
\eeq
Finally, substituting \eq{hn} into \eq{image} and \eq{cdx} we find
\beql{cdt}
\frac{d\sigma_\text{cd}^{\dip+A}}{dt}=\frac{1}{16\pi} \left| 
2\int d^2b \left( 1-e^{-\frac{1}{2} \sigma_\text{tot}^{\dip+N}(s)\rho T_A(b)}\right) e^{i\b b\cdot \b \Delta}
\right|^2\,.
\eeq

To estimate the $t$-dependence of the coherent cross section  we can use a simple model for the $b$-distribution. Denote $\frac{1}{2} \sigma_\text{tot}^{\dip+N}(s)\rho T_A(b)=\Omega S(b)$ and let the profile function $S(b)$ be given by the step function $S(b)= \theta(R_A-b)$. Neglecting contribution of the  diffuse region at the nucleus edge is a reasonable approximation in the case of coherent diffraction because the main contribution stems from $b< R_A$ impact parameters. Substituting into \eq{cdt} and \eq{cdx} we get the well-known result (see e.g.\ \cite{Levin:1997mg}) 
\beql{coh-t}
\frac{d\sigma_{cd}^{\dip+A}}{dt}\frac{1}{\sigma_{cd}^{\dip+A}}= \frac{J_1^2(R_A\sqrt{-t})}{|t|}\,.
\eeq
Because \eq{coh-t} does not depend on $\Omega$ this formula also gives $t$-dependence of the diffractive coherent gluon production:
\beql{incl-coh-t}
\frac{d\sigma_\text{cd}^{\gamma^*A}(Q^2,x, k,y)}{d^2 k\, dy\,dt} =\frac{J_1^2(R_A\sqrt{-t})}{|t|}\,\frac{d\sigma_\text{cd}^{\gamma^*A}(Q^2,x, k,y)}{d^2 k\, dy}\,.
\eeq

\subsection{Incoherent diffraction}

 Coherent diffraction includes only events in which nucleus stays intact. However, generally the nucleus can be excited and subsequently  decays into colorless remnants. Total diffractive cross section of this process is given by 
 \beql{tot.dif}
\sigma_\text{dif}^{\dip+A}=\int d^2b\, \aver{\left| \Gamma^{\dip+ A}(s,\b b)\right|^2}\,.
\eeq
The difference between \eq{tot.dif} and \eq{cdx} measures dispersion of the scattering amplitude in the impact parameter space. The corresponding physical process is a part of inelastic cross section and is called incoherent diffraction: 
 \beql{idx}
\sigma_\text{id}^{\dip+ A}=\int d^2b\, \aver{\left| \Gamma^{\dip+ A}(s,\b b)\right|^2}-\left| \aver{\Gamma^{\dip+ A}(s,\b b)}\right|^2\,.
\eeq
Clearly, the incoherent diffraction stems from the region near the nucleus edge (`diffuse region'). Indeed, at $b\ll R_A$ all partial dipole-nucleon amplitudes are close to the black disk limit, while at $b\gg R_A$ they all vanish. 

To derive the $t-$dependence of the incoherent diffraction cross section we define similarly to 
\eq{image}
\beq\label{image2}
\Gamma^{\dip+ A}(s,\b \Delta,\{ \b b_a\})= 2\int d^2b \,\Gamma^{\dip+ A}(s,\b b, \{\b b_a\}) e^{i\b b\cdot \b \Delta}\,.
\eeq
Then \eq{tot.dif} reads:
\begin{eqnarray}\label{f.tot}
\frac{d\sigma_\text{dif}}{dt}&=& \frac{1}{16\pi}\aver{\left| \Gamma^{\dip+ A}(s,\b \Delta,\{ \b b_a\})\right|^2}\\
&=& \frac{1}{4\pi}\int d^2b \int d^2b' e^{i\b \Delta\cdot (\b b-\b b')}\aver{
\left[1-\prod _{a=1}^A\left(
1-\Gamma^{\dip+ N}(s,\b b-\b b_a)\right)\right]
\left[1-\prod _{a=1}^A\left(
1-\Gamma^{\dip+ N}(s,\b b'-\b b_a)\right)\right]^\dagger
}\nonumber\\
&=& \frac{1}{4\pi}\int d^2b \int d^2b' e^{i\b \Delta\cdot (\b b-\b b')} \left[ 
1-e^{-\sum_a \aver{\Gamma^{\dip+ N}(s,\b b-\b b_a)}}-e^{-\sum_a \aver{\Gamma^{\dip+ N}(s,\b b'-\b b_a)}}\right. \nonumber\\
&& \left. + e^{-\sum_a \aver{\Gamma^{\dip+ N}(s,\b b-\b b_a)}+\sum_a \aver{\Gamma^{\dip+ N}(s,\b b'-\b b_a)}-\aver{\Gamma^{\dip+ N}(s,\b b-\b b_a)\Gamma^{\dip+ N}(s,\b b'-\b b_a)}}
\right]\,.
\end{eqnarray}
Upon subtracting the coherent diffraction part
\beql{cdp}
\frac{d\sigma_\text{cd}}{dt}=\frac{1}{4\pi}\int d^2b \int d^2b' e^{i\b \Delta\cdot (\b b-\b b')}
\left( 1- e^{-\sum_a \aver{\Gamma^{\dip+ N}(s,\b b-\b b_a)}}\right)
\left( 1- e^{-\sum_a \aver{\Gamma^{\dip+ N}(s,\b b'-\b b_a)}}\right)
\eeq
we end up with 
\begin{eqnarray}\label{idp}
\frac{d\sigma_\text{id}}{dt}&=&\frac{1}{4\pi}\int d^2b \int d^2b' e^{i\b \Delta\cdot (\b b-\b b')}
\left[ 1-e^{-\sum_a\aver{\Gamma^{\dip+ N}(s,\b b-\b b_a)\Gamma^{\dip+ N}(s,\b b'-\b b_a)}}\right]\nonumber\\
&&
\times e^{-\sum_a \left[\aver{\Gamma^{\dip+ N}(s,\b b-\b b_a)}+ \aver{\Gamma^{\dip+ N}(s,\b b'-\b b_a)}-\aver{\Gamma^{\dip+ N}(s,\b b-\b b_a)\Gamma^{\dip+ N}(s,\b b'-\b b_a)}\right]}\,.
\end{eqnarray}
Since elastic $q\bar q N$ cross section is small compared with the inelastic one (as it contains extra $\as^2$ \cite{Tuchin:2008np}), we expand \eq{idp}:
\begin{eqnarray}
\frac{d\sigma_\text{id}}{dt}&=&\frac{1}{4\pi}\int d^2b \int d^2b' e^{i\b \Delta\cdot (\b b-\b b')}
e^{-\sum_a \left[\aver{\Gamma^{\dip+ N}(s,\b b-\b b_a)}+ \aver{\Gamma^{\dip+ N}(s,\b b'-\b b_a)}\right]}\nonumber\\
&&\times \sum_a\aver{\Gamma^{\dip+ N}(s,\b b-\b b_a)\Gamma^{\dip+ N}(s,\b b'-\b b_a)}\label{idp2}
\\
&=& \frac{1}{4\pi}\int d^2b_a \left| \int d^2b\, e^{i\b \Delta\cdot \b b} e^{-\rho T_A(b) \Gamma^{\dip+ N}(s,0)}\Gamma^{\dip+ N}(\b b-\b b_a)\right|^2 \rho T_A(\b b_a)\label{idp3}
\end{eqnarray}
 Because $|\b b-\b b_a|\sim R_p\ll b_a\sim R_A$, \eq{idp3} becomes 
 \beql{idp4}
 \frac{d\sigma_\text{id}}{dt}=\frac{1}{4\pi}\int d^2b_a\,e^{-2\rho T_A(b_a) \Gamma^{\dip+ N}(s,0)} \left| \int d^2b\, e^{i\b \Delta\cdot \b b} \Gamma^{\dip+ N}(\b b)\right|^2 \rho T_A(\b b_a)\,.
 \eeq
Finally, using \eq{dN} we derive the desired result
\beql{idp5}
 \frac{d\sigma_\text{id}}{dt}=\frac{1}{4\pi}\frac{\sigma^{\dip+ N}_\text{tot}(s)}{2} e^{-\frac{1}{2}tR_p^2}\int d^2b_a\,e^{-2\rho T_A(b_a) \Gamma^{\dip+ N}(s,0)}  \rho T_A(\b b_a) = \frac{R_p^2}{2} e^{-\frac{1}{4}|t|R_p^2}\,\sigma_\text{id}\,. 
 \eeq
 As in the case of coherent diffraction, $t$-dependence of the cross section for incoherent diffraction is independent of other kinematic variables and therefore \eq{idp5} describes also $t$-dependence of incoherent diffractive hadron production:
 \beql{incl-incoh-t}
\frac{d\sigma_\text{id}^{\gamma^*A}(Q^2,x, k,y)}{d^2 k\, dy\,dt} =\frac{d\sigma_\text{id}^{\gamma^*A}(Q^2,x, k,y)}{d^2 k\, dy}\,\frac{R_p^2}{2} e^{-\frac{1}{4}|t|R_p^2}\,.
\eeq

\begin{figure}[ht]
\begin{tabular}{cc}
      \includegraphics[height=5.4cm]{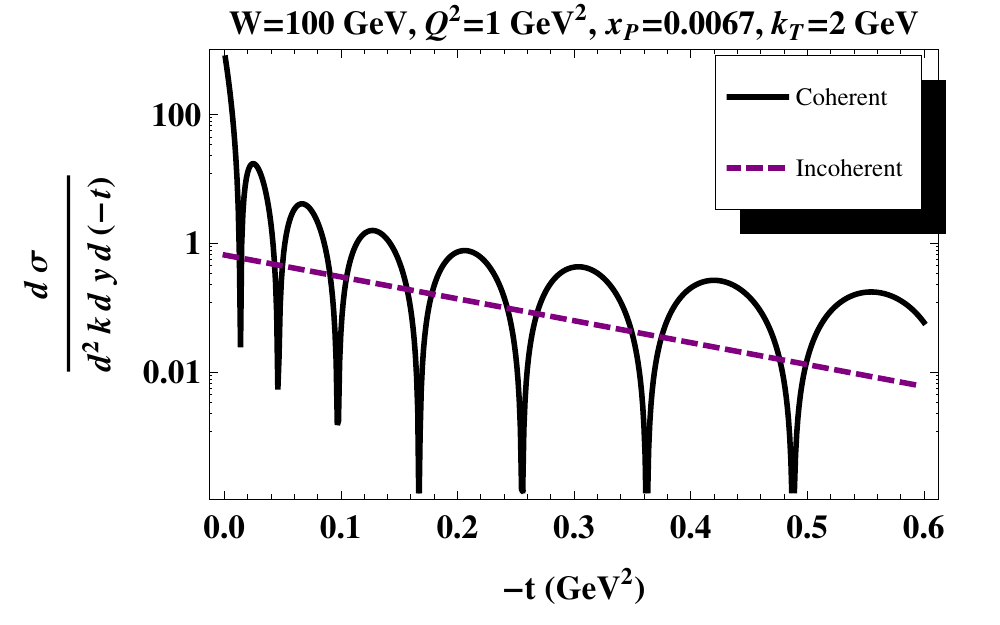} &
      \includegraphics[height=6.1cm]{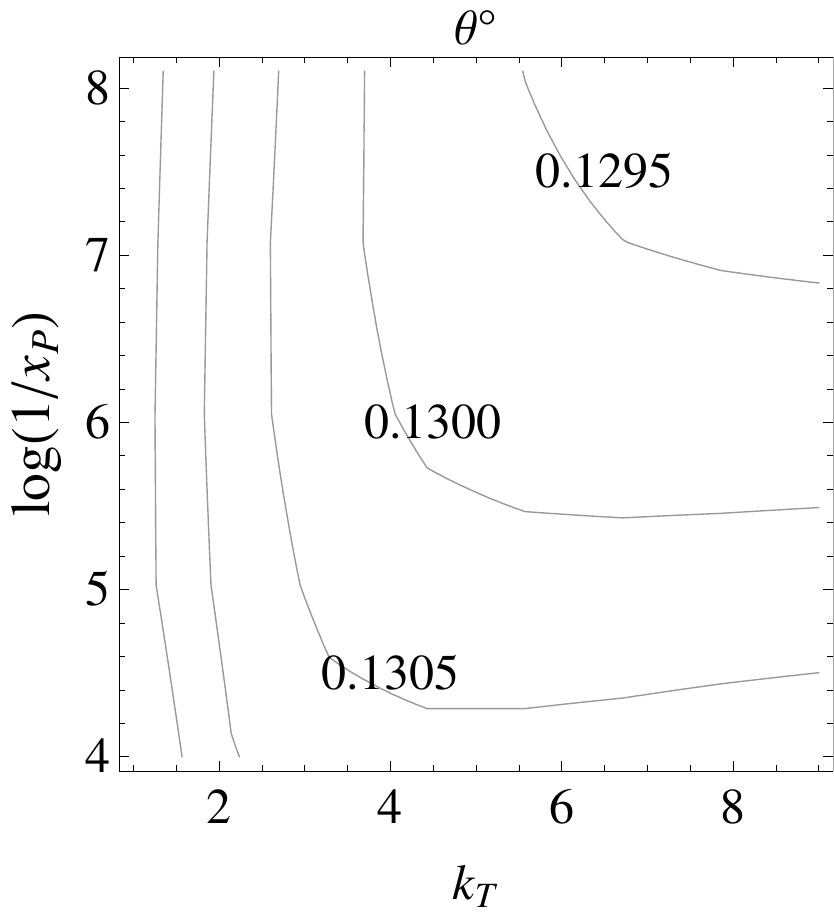}\\
      $(a)$ & $(b)$ 
      \end{tabular}
  \caption{ (a) $t$-dependence of diffractive coherent (solid line) and incoherent (broken line) hadron production. (b) Contour plot of the scattering angle $\theta'$ at which coherent and incoherent diffractive hadron production cross sections are equal at given $\b k$ and $y=\ln(1/\xP)$.}
\label{fig:t-dep}
\end{figure}
$t$-dependence of diffractive coherent and incoherent hadron production is shown in \fig{fig:t-dep}.  
In \fig{fig:t-dep}(a) we compare the coherent and incoherent diffractive production as a function of the transferred momentum $t$ for a particular choice of the collision kinematics. We observe the following general features: {\sl (i)} At small $|t|$ coherent diffraction dominates over incoherent one by several orders of magnitude; {\sl (ii)} At $|t|\approx R_p^{-2}$ coherent and incoherent contributions coincide, we will call the corresponding scattering angle $\theta'$; at $|t|>R_p^{-2}$ the two contributions are similar on average; {\sl (iii)}  Features {\sl (i)} and {\sl (ii)} hold for a wide range of parameters. In particular, $\theta'$ is nearly constant $\theta'\approx 0.13^\text{o}$ as is seen in \fig{fig:t-dep}(b). Scattering angles of this size may be experimentally accessible at EIC.

\section{Summary}

In this paper we discussed coherent and incoherent diffractive gluon production in DIS off heavy nuclei in the proposed kinematic region of Electron Ion Collider.  Our approach is based on the dipole model introduced in \cite{Mueller:1989st}. It allows representing cross sections for high energy hadronic scattering as a convolution of hadronic light-cone wave-functions with the multipole scattering amplitudes. In our case, virtual photon wave function is determined by the perturbative QED and is given by \eq{w-f}. Dipole-nucleus interaction can in turn be represented as a product of  dipole density \eq{ndiff} in transverse coordinate space, satisfying the BFKL equation \cite{Kuraev:1977fs,Balitsky:1978ic}, and the dipole-nucleus forward elastic scattering amplitude as displayed in  \eq{cd10},\eq{all2}, satisfying QCD evolution equations in the low $x$ region \cite{Balitsky:1995ub,Kovchegov:1999yj}. These formulas are derived in the leading logarithmic approximation $\as \ln(1/x)\sim 1$, $\as\ln(1/\beta)\sim 1$, which defines the kinematic region where the results of our calculations are applicable. Note, that  hard perturbative factorization is generally broken at low $x$, because scattering in this region is characterized by small longitudinal momentum transfer (see e.g.\ \cite{Tuchin:2010pv}). At moderate $x$ and large $Q^2$, our formulas reduce to the leading order hard perturbative QCD expressions that can be cast in the factorized form using the diffractive parton distributions  \cite{Trentadue:1993ka,Berera:1995fj,Collins:1997sr,Hautmann:1998xn}. 

The main results of our calculations are displayed in \fig{main-figure} and \fig{fig:t-dep}. We found that nuclear modification factor strongly varies with nuclear weight, and the functional dependence on $A$ is qualitatively different for coherent and incoherent processes. Similarly to diffractive hadron production in $pA$ collisions \cite{Tuchin:2008np}, nuclear effects in coherent diffractive DIS is strongly dependent on rapidity of produced hadron, whereas they are almost absent in the case of incoherent diffraction. We also made a peculiar observation that the  nuclear modification factor for both diffractive channels is essentially independent of the photon virtuality in the region $1<Q^2<25$~GeV$^2$. Finally, our study of non-forward diffractive hadron production indicates feasibility of experimentally separation of 
coherent and incoherent diffractive contributions at EIC.

\acknowledgments
This work  was supported in part by the U.S. Department of Energy under Grant No.\ DE-FG02-87ER40371.



\begin{thebibliography}{80}

\bibitem{Levin:1997mg}
  E.~Levin,
  arXiv:hep-ph/9710546.



\bibitem{Gribov:1983tu}
L.~V.~Gribov, E.~M.~Levin, and M.~G.~Ryskin,
Phys.\ Rept.\  {\bf 100}, 1 (1983).

\bibitem{Mueller:1986wy}
A.~H.~Mueller and J.~W.~Qiu,
Nucl.\ Phys.\ B {\bf 268}, 427 (1986).

  


\bibitem{McLerran:1993ni}
L.~D.~McLerran and R.~Venugopalan,
Phys.\ Rev.\ D {\bf 49}, 2233 (1994)
[arXiv:hep-ph/9309289];
Phys.\ Rev.\ D {\bf 49}, 3352 (1994)
[arXiv:hep-ph/9311205];
Phys.\ Rev.\ D {\bf 50}, 2225 (1994)
[arXiv:hep-ph/9402335].

\bibitem{Kovchegov:1996ty}
  Y.~V.~Kovchegov,
  Phys.\ Rev.\  D {\bf 54}, 5463 (1996)
  [arXiv:hep-ph/9605446];
  Phys.\ Rev.\  D {\bf 55}, 5445 (1997)
  [arXiv:hep-ph/9701229].



\bibitem{Balitsky:1995ub}
  I.~Balitsky,
  Nucl.\ Phys.\  B {\bf 463}, 99 (1996)
  [arXiv:hep-ph/9509348];
  I.~Balitsky,
  Phys.\ Rev.\ Lett.\  {\bf 81}, 2024 (1998)
  [arXiv:hep-ph/9807434];
  I.~Balitsky,
  Phys.\ Rev.\  D {\bf 60}, 014020 (1999)
  [arXiv:hep-ph/9812311].



\bibitem{JalilianMarian:1997jx}
  J.~Jalilian-Marian, A.~Kovner, A.~Leonidov and H.~Weigert,
  Nucl.\ Phys.\  B {\bf 504}, 415 (1997)
  [arXiv:hep-ph/9701284];
  J.~Jalilian-Marian, A.~Kovner, A.~Leonidov and H.~Weigert,
  Phys.\ Rev.\  D {\bf 59}, 014014 (1999)
  [arXiv:hep-ph/9706377].

\bibitem{Jalilian-Marian:1997dw}
J.~Jalilian-Marian, A.~Kovner, and H.~Weigert,
Phys.\ Rev.\ D {\bf 59}, 014015 (1999)
[arXiv:hep-ph/9709432];

\bibitem{Kovner:2000pt}
A.~Kovner, J.~G.~Milhano, and H.~Weigert,
Phys.\ Rev.\ D {\bf 62}, 114005 (2000)
[arXiv:hep-ph/0004014];
H.~Weigert,
Nucl.\ Phys.\ A {\bf 703}, 823 (2002)
[arXiv:hep-ph/0004044].



\bibitem{Iancu:2000hn}
  E.~Iancu, A.~Leonidov and L.~D.~McLerran,
  Nucl.\ Phys.\  A {\bf 692}, 583 (2001)
  [arXiv:hep-ph/0011241].

\bibitem{Ferreiro:2001qy}
  E.~Ferreiro, E.~Iancu, A.~Leonidov and L.~McLerran,
  Nucl.\ Phys.\  A {\bf 703}, 489 (2002)
  [arXiv:hep-ph/0109115].

\bibitem{Gotsman:1996ix}
  E.~Gotsman, E.~Levin and U.~Maor,
  Nucl.\ Phys.\  B {\bf 493}, 354 (1997)
  [arXiv:hep-ph/9606280].

      \bibitem{Bartels:1998ea}
        J.~Bartels, J.~R.~Ellis, H.~Kowalski and M.~Wusthoff,
        Eur.\ Phys.\ J.\  C {\bf 7}, 443 (1999)
        [arXiv:hep-ph/9803497].


  
\bibitem{GolecBiernat:1998js}
  K.~J.~Golec-Biernat and M.~Wusthoff,
  Phys.\ Rev.\  D {\bf 59}, 014017 (1999)
  [arXiv:hep-ph/9807513].

\bibitem{GolecBiernat:1999qd}
  K.~J.~Golec-Biernat and M.~Wusthoff,
  Phys.\ Rev.\  D {\bf 60}, 114023 (1999)
  [arXiv:hep-ph/9903358].



\bibitem{Gotsman:2000gb}
  E.~Gotsman, E.~Levin, M.~Lublinsky, U.~Maor, and K.~Tuchin,
  arXiv:hep-ph/0007261.

\bibitem{GolecBiernat:2005fe}
  K.~J.~Golec-Biernat and C.~Marquet,
  Phys.\ Rev.\  D {\bf 71}, 114005 (2005)
  [arXiv:hep-ph/0504214].
  
\bibitem{Li:2008bm}
  Y.~Li and K.~Tuchin,
  arXiv:0802.2954 [hep-ph].

\bibitem{Kovchegov:2001ni}
  Y.~V.~Kovchegov,
  Phys.\ Rev.\  D {\bf 64}, 114016 (2001)
  [Erratum-ibid.\  D {\bf 68}, 039901 (2003)]
  [arXiv:hep-ph/0107256].

\bibitem{Kovner:2001vi}
  A.~Kovner and U.~A.~Wiedemann,
  Phys.\ Rev.\  D {\bf 64}, 114002 (2001)
  [arXiv:hep-ph/0106240].
  
\bibitem{Kovner:2006ge}
  A.~Kovner, M.~Lublinsky, and H.~Weigert,
  Phys.\ Rev.\  D {\bf 74}, 114023 (2006)
  [arXiv:hep-ph/0608258].



\bibitem{Wusthoff:1997fz}
  M.~Wusthoff,
  Phys.\ Rev.\  D {\bf 56}, 4311 (1997)
  [arXiv:hep-ph/9702201].


\bibitem{Marquet:2004xa}
  C.~Marquet,
  Nucl.\ Phys.\  B {\bf 705}, 319 (2005)
  [arXiv:hep-ph/0409023].

\bibitem{Marquet:2007nf}
  C.~Marquet,
  Phys.\ Rev.\  D {\bf 76}, 094017 (2007)
  [arXiv:0706.2682 [hep-ph]].

\bibitem{Munier:2003zb}
  S.~Munier and A.~Shoshi,
  Phys.\ Rev.\  D {\bf 69}, 074022 (2004)
  [arXiv:hep-ph/0312022].


\bibitem{Kugeratski:2005ck}
  M.~S.~Kugeratski, V.~P.~Goncalves and F.~S.~Navarra,
  Eur.\ Phys.\ J.\  C {\bf 46}, 413 (2006)
  [arXiv:hep-ph/0511224].



\bibitem{Kowalski:2007rw}
  H.~Kowalski, T.~Lappi and R.~Venugopalan,
  Phys.\ Rev.\ Lett.\  {\bf 100}, 022303 (2008)
  [arXiv:0705.3047 [hep-ph]].

\bibitem{Kowalski:2008sa}
  H.~Kowalski, T.~Lappi, C.~Marquet and R.~Venugopalan,
  Phys.\ Rev.\  C {\bf 78}, 045201 (2008)
  [arXiv:0805.4071 [hep-ph]].


\bibitem{Li:2008jz}
  Y.~Li and K.~Tuchin,
  arXiv:0803.1608 [hep-ph].
  

  
\bibitem{Li:2008se}
  Y.~Li and K.~Tuchin,
  Phys.\ Rev.\  C {\bf 78}, 024905 (2008)
  [arXiv:0806.2087 [hep-ph]].

  
\bibitem{Tuchin:2008np}
  K.~Tuchin,
  Phys.\ Rev.\  C {\bf 79}, 055206 (2009)
  [arXiv:0812.1519 [hep-ph]].

  

\bibitem{Kuraev:1977fs}
  E.~A.~Kuraev, L.~N.~Lipatov, and V.~S.~Fadin,
  Sov.\ Phys.\ JETP {\bf 45}, 199 (1977)
  [Zh.\ Eksp.\ Teor.\ Fiz.\  {\bf 72}, 377 (1977)].

\bibitem{Balitsky:1978ic}
  I.~I.~Balitsky and L.~N.~Lipatov,
  Sov.\ J.\ Nucl.\ Phys.\  {\bf 28} (1978) 822
  [Yad.\ Fiz.\  {\bf 28} (1978) 1597].





\bibitem{Kniehl:2000hk}
  B.~A.~Kniehl, G.~Kramer and B.~Potter,
  Nucl.\ Phys.\  B {\bf 597}, 337 (2001)
  [arXiv:hep-ph/0011155].




\bibitem{Kowalski:2006hc}
  H.~Kowalski, L.~Motyka and G.~Watt,
  Phys.\ Rev.\  D {\bf 74}, 074016 (2006)
  [arXiv:hep-ph/0606272].

\bibitem{Iancu:2002tr}
  E.~Iancu, K.~Itakura and L.~McLerran,
  Nucl.\ Phys.\  A {\bf 708}, 327 (2002)
  [arXiv:hep-ph/0203137].

\bibitem{Gotsman:2002zi}
  E.~Gotsman, E.~Levin, M.~Lublinsky, U.~Maor and K.~Tuchin,
  Nucl.\ Phys.\  A {\bf 697} (2002) 521.


\bibitem{Mueller:1989st}
  A.~H.~Mueller,
  Nucl.\ Phys.\  B {\bf 335}, 115 (1990).


\bibitem{Kovchegov:1999yj}
  Y.~V.~Kovchegov,
  Phys.\ Rev.\  D {\bf 60}, 034008 (1999)
  [arXiv:hep-ph/9901281].

\bibitem{Tuchin:2010pv}
  K.~Tuchin,
  arXiv:1012.4212 [hep-ph].

\bibitem{Trentadue:1993ka}
  L.~Trentadue and G.~Veneziano,
  Phys.\ Lett.\  B {\bf 323}, 201 (1994).

\bibitem{Berera:1995fj}
  A.~Berera and D.~E.~Soper,
  Phys.\ Rev.\  D {\bf 53}, 6162 (1996)
  [arXiv:hep-ph/9509239].

\bibitem{Collins:1997sr}
  J.~C.~Collins,
  Phys.\ Rev.\  D {\bf 57}, 3051 (1998)
  [Erratum-ibid.\  D {\bf 61}, 019902 (2000)]
  [arXiv:hep-ph/9709499].


\bibitem{Hautmann:1998xn}
  F.~Hautmann, Z.~Kunszt and D.~E.~Soper,
  Phys.\ Rev.\ Lett.\  {\bf 81}, 3333 (1998)
  [arXiv:hep-ph/9806298].

\end{thebibliography}
\end{document}